\documentclass{easychair}

\usepackage{doc}

\def\*#1{\mathbf{#1}}
\usepackage{hyperref}       % hyperlinks

\usepackage{url}            % simple URL typesetting
\usepackage{booktabs}       % professional-quality tables
\usepackage{amsfonts}       % blackboard math symbols
\usepackage{nicefrac}       % compact symbols for 1/2, etc.
\usepackage{microtype}      % microtypography
\usepackage{mathtools}
\usepackage{wrapfig}

\usepackage{tikz}
\usepackage{subfig}

\newcommand{\linfour}{{\sf LinProx\/}}
\newcommand{\linsix}{{\sf LinProxTh\/}}
\newcommand{\nonlinfour}{{\sf NLinProx\/}}
\newcommand{\nonlinsix}{{\sf NLinProxTh\/}}
\newcommand{\sdvtool}{{\sf SDVTool\/}}

\title{Verifying safety of an autonomous spacecraft rendezvous mission (Experience Report)}
\author{
Nicole Chan and Sayan Mitra
}
\institute{
  Coordinated Science Laboratory,\\
  University of Illinois at Urbana-Champaign\\
  \email{\{nschan3,mitras\}@illinois.edu}
}

\authorrunning{N. Chan and S. Mitra}
\titlerunning{Verification of Autonomous Spacecraft Rendezvous}

\begin{document}
\maketitle
%% NOTES TO SELF:
%% Mainly an experience report on C2E2. But also compare with some SpaceEx results. And uses a prototype MATLAB program.
%% However also presents a benchmark problem (the space problem already presented as benchmark by Scott) with passive safety --> is the passive mode stiff? Clearly the method of obtaining reachsets can differ greatly. Should compare SpaceEx results between LGG and Phaver algorithms.
%% Also loosely a tool paper if you include the MATLAB stuff that we want to add into C2E2 and improvements in C2E2 (can handle nonlinear behaviors beyond polynomials). Can't recall if/what C2E2 does for non-deterministic transitions (ask Chuchu).
%% Note extra overapproximation (and perhaps excessive error growth) from collecting states over non-deterministic transitions. We're technically supposed to account for the range of time from the collected transition states and propagate the interval throughout all subsequent reachtubes. Fortunately the system is time-invariant so the reachtubes will not change depending on "start time" but because a future transition (to passive) IS time dependent, we should "grow" the sets that may transition to account for this time shift. [Maybe don't bring this up at all but keep in mind and fix]

\begin{abstract}
%
% Autonomous space operations have demonstrated the need for rigorous safety guarantees. A fundamental orbital maneuver is known as rendezvous, 
A fundamental maneuver in autonomous space operations is known as rendezvous, where a spacecraft navigates to and approaches another spacecraft. In this case study, we present linear and nonlinear benchmark models of an active chaser spacecraft performing rendezvous toward a passive, orbiting target. The system is modeled as a hybrid automaton, where the chaser must adhere to different sets of constraints in each discrete mode. A switched LQR controller is designed accordingly to meet this collection of physical and geometric safety constraints, while maintaining liveness in navigating toward the target spacecraft.
We extend this benchmark problem to check for passive safety, which is collision avoidance along a passive, propulsion-free trajectory that may be followed in the event of system failures.
We show that existing hybrid verification tools like  SpaceEx, C2E2, and our own implementation of a simulation-driven verification tool can robustly verify this system with respect to the requirements, and a variety of relevant initial conditions.
%The case studies also point to weaknesses of existing tools in handling ill-conditioned models. 

\end{abstract}

\section{Introduction}
\label{sec:intro}

A new age of deep space exploration is underway with several ongoing public-private partnerships. A groundwork for a possible mission to Mars in the 2030s is also underway~\cite{obama_mars}.
Autonomous operations where a spacecraft can operate independent of human control in a wide variety of conditions are essential for  deployment, construction, and maintenance missions in space. 
Despite many spectacular successes like the  Mars landing of the Curiosity rover, ensuring safety of autonomous spacecraft operations remains a daunting challenge. The cost of failures can be extreme. 
For example, NASA's DART spacecraft was designed to rendezvous with the MUBLCOM satellite. In 2005, approximately 11 hours into a 24-hour mission, DART's propellant supply depleted due to excessive use of thrusters, and it began maneuvers for departure. In the process it collided with MUBLCOM; it met only 11 of 27 mission objectives, and the failure resulted in a loss exceeding \$1 million. 
In another incident, a navigation error caused the Mars Climate Orbiter to reach as low as 57 kilometers, where it was intended to enter an orbit at an altitude of 140-150 kilometers. The spacecraft was destroyed by the resulting atmospheric stresses and friction and the cost incurred was \$85 million. 
%The error resulted from inconsistent units between navigational data and the on-board software because the software was adapted from an earlier project and not adequately tested on the new spacecraft. 
These incidents, and several others in~\cite{wong:sw} highlight the consequences of failures in space applications and demonstrate the need for more rigorous testing before deployment.

Although formal verification has played an important role in design and safety analysis of spacecraft hardware and software (see, for example~\cite{Holzmann:2014:MC} and the references therein), they have not been used for model-based design and system-level verification and validation.
% This is in part because the hybrid system verification tools were not mature enough to handle spacecraft dynamics.  
%
In this paper, we present and verify a realistic and challenging spacecraft maneuver problem called autonomous rendezvous, proximity operations, and docking (ARPOD). 
The original hybrid control design problem and its variants are introduced by Jewison and Erwin in~\cite{erwin-cdc16}. 
ARPOD is a fundamental set of operations for a variety of space missions such as on-orbit transfer of personnel~\cite{Woffinden:rendezvous}, resupply for on-orbit personnel~\cite{Pinard2007}, assembly~\cite{Zimpfer:rendezvous}, 
servicing, repair, and refueling~\cite{Galabova2003}.

The basic setup for ARPOD consists of a passive module or a {\em target\/}  (launched separately into orbit) and a {\em chaser\/} spacecraft that must transport the passive module to an on-orbit assembly location. The chaser maintains a relative bearing measurement to the target, but initially it may be too far away from the target to use its range sensors. Range measurements become available within a given range, giving the chaser accurate relative positioning data so that it can stage itself to dock the target. The target must be docked with a specific angle of approach and closing velocity, so as to avoid collision and ensure that the docking mechanisms on each body will mate. Furthermore, the docking procedure must be completed before the chaser goes into eclipse and loses vision-based sensor data. 
Finally, it is necessary for the system to ensure {\em passive safety\/}. That is, the chaser spacecraft should maintain safe separation from the target even if it loses power and communication during its mission. 

In this paper, we present a suite of hybrid models for the rendezvous portion of the ARPOD mission that can serve as benchmarks for verification tools and serve as building-blocks for more complex operations. 
We present nonlinear models that consist of nonlinear orbital dynamics (\nonlinfour{} and \nonlinsix{}) and linearized models (\linfour{} and \linsix{}) using the Clohessy-Wiltshire-Hill (CWH) equations~\cite{1960}. %that are commonly used in astrodynamics. 
The rendezvous operation is further subdivided into two phases: Proximity Operations A and B, such that phase A captures an interval of larger ranges than the interval of ranges in phase B. In other words, the chaser enters phase A first and as it moves closer to the target, it enters phase B. After this two-phase rendezvous, the chaser enters a docking phase/maneuver (i.e. when the chaser is less than 10 meters from its target). We disregard the requirements for this phase for this paper. We develop a switched state-feedback controller using Linear Quadratic Regulation (LQR) for the rendezvous phases. 
%
%The  models also include all the safety requirements for avoiding collisions, maintaining line-of-sight (LOS) and  velocity constraints, and obeying physical limitations on propulsion. 
%
The position-triggered transitions brought about by the switching controller are urgent, resulting in a deterministic hybrid automaton. However, we extend the ARPOD problem in~\cite{erwin-cdc16} to include the passive safety requirement by introducing a nondeterministic time-triggered transition to the passive mode.
%
%
%Finally, we present versions of the  models that approximate the nonlinear guards and invariants with polytopes, (making it amenable for analysis with existing hybrid verification tools).

We have successfully verified the requirements for most of the models using existing hybrid verification tools SpaceEx~\cite{DBLP:conf/cav/FrehseGDCRLRGDM11} and C2E2~\cite{DuggiralaMVP15,FanQM0D16:CAV},
and also our new MatLab implementation of a simulation-driven verification algorithm (\sdvtool{}) for linear hybrid models. \sdvtool{} improves C2E2's reachability algorithm, with a new technique~\cite{Duggirala2016} for obtaining reachsets for linear systems.
We obtain verification results for an array of varied initial state configurations and passive transitions times to show the robustness and limits of the switched LQR controller. 
%We show that the verification results are robust to changes in the initial configurations and passive transition times. 
%
The experiments, in particular on the passive safety requirement, have demonstrated a weakness in simulation-driven verification approaches in handling ill-conditioned models which suggest a need for further research.
Overall, we believe that our results and approaches establish feasibility of system-level verification of autonomous space operations, and they provide a foundation for the analysis of more sophisticated maneuvers in the future.

\section{Related work}
\label{sec:related}

There are few academic works on system-level verification of autonomous spacecraft. 
A survey of general verification approaches and how they may apply to small satellite  systems is presented in~\cite{nasa_survey}.
% this paper: %https://ti.arc.nasa.gov/publications/23631/download/
%
 Architecture and Analysis Design Language (AADL) and verification and validation (V\&V)  over AADL models for satellite systems have been reported in~\cite{bozzano2010formal}
% this paper:
%
% http://web1.see.asso.fr/erts2010/Site/0ANDGY78/Fichier/PAPIERS%20ERTS%202010%202/ERTS2010_0098_final.pdf

An feasibility study for applying formal verification of 
autonomous satellite maneuvers is presented in~\cite{JGMDE:satellite2012}.
That approach relied on creating rectangular  abstractions (dynamics of the form $\dot x \in [a,b]$) of the satellites dynamics through hybridization and verification using PHAVer~\cite{Frehse:hscc05} and SpaceEx~\cite{DBLP:conf/cav/FrehseGDCRLRGDM11}. 
The generated abstract models have simple dynamics but hundreds of locations, and also, the analysis is necessarily conservative. In contrast, the approaches presented in this paper work directly with the linear (nonlinear) hybrid dynamics.

The ARPOD challenge~\cite{erwin-cdc16} has been taken up by several researchers in proposing a variety of control strategies. 
A two-stage optimal control strategy is developed in~\cite{Farahani-cdc16}, where the first part involves trajectory planning under a differentially-flat system and the second part implements Model Predictive Control on a linearized model. 
A supervisor is introduced to robustly coordinate a family of hybrid controllers in~\cite{sanfelice2016cdc}. 
Safe reachsets (i.e. ReachAvoid sets) are computed for the ARPOD mission in~\cite{oishi2016cdc} and used to solve for minimum fuel and minimum time trajectories. 
%Previous FM paper by Taylor, Sayan et al. and the citation there in.
%http://link.springer.com/chapter/10.1007%2F978-3-642-32759-9_22#page-1

\section{Spacecraft Rendezvous Model}
\label{sec:model}

In this section, we present the detailed development of the hybrid models. First we present the orbital dynamics of the spacecraft in Sections~\ref{ssec:dynamics}-\ref{ssec:lindynamics}. Then in Sections~\ref{sec:hybrid-control}-\ref{sec:lqr} we present a hybrid controller. Finally, we state the various mission constraints in Section~\ref{sec:safeSets}. 

%%%%%%%%%%%%%%%%%%%%%%%%%%%
\subsection{Nonlinear relative motion dynamics}
\label{ssec:dynamics}

The dynamics of the two spacecraft in orbit---the {\em target\/} and the {\em chaser\/}---are derived from Kepler's laws. We use the simplest case for relative motion in space, where the two spacecraft are restricted to the same orbital plane, resulting in two-dimensional, planar motion. The so called Hill's relative coordinate frame is used. As shown in Figure~\ref{Hill}, Hill's frame is centered on the target spacecraft, with $+\hat{\*i}$-direction pointing radially outward from the Earth, $+\hat{\*k}$-direction normal from the orbital plane, and $+\hat{\*j}$-direction completing a right-handed system. We further assume that  the target moves on a circular orbit, and thus, the $\hat{\*j}$-direction aligns with the tangential velocity of the target. 

The restriction on the target's orbit implies that the target-centered frame rotates with constant angular velocity. We will assume the target is in geostationary equatorial orbit (GEO), so its angular velocity is $n = \sqrt{\frac{\mu}{r^3}}$, where $\mu=3.698\times 10^{14} m^3/s^2$ and $r=42164 km$. The chaser's position is represented by the vector $x \*i + y\*j $, and the chaser's thrusters provide acceleration in the form of $F_x\*i + F_y\*j$. The following equations are derived using Kepler's laws and constitute the nonlinear model of the spacecraft dynamics.
\begin{align}
\begin{split}
\ddot{x} &= n^2 x + 2n\dot{y} + \frac{\mu}{r^2} - \frac{\mu}{r_c^3}(r+x) + \frac{F_x}{m_c}, \\ 
\ddot{y} &= n^2 y - 2n\dot{x} - \frac{\mu}{r_c^3}y + \frac{F_y}{m_c}, 
\end{split}
\label{eq:nonlin1}
\end{align}
where $r_c = \sqrt{(r+x)^2 + y^2}$ is the distance between the chaser and Earth and $m_c = 500$kg is the mass of the chaser.

%%%%%%%%%%%%%%%%%%%%%%%%%%%
\subsection{Linear dynamics}
\label{ssec:lindynamics}
Linearization of these equations about the system's equilibrium
 point results in the Clohessy-Wiltshire-Hill (CWH) equations \cite{1960}, which are commonly used to capture the relative motion dynamics of two satellites within a reasonably close range. These equations are:
\begin{align}
\begin{split}
\ddot{x} &= 3n^2 x + 2n\dot{y} + \frac{F_x}{m_c}, \\
\ddot{y} &= - 2n\dot{x} + \frac{F_y}{m_c}. 
\end{split}
\label{eq:lin1}
\end{align}

Let the state vector be denoted by $\vec{x} = [x, y, \dot{x}, \dot{y}]^{T}$. The state-space form of these linear time-invariant (LTI) equations is:
$$\dot{\vec{x}} = A\vec{x} + B\vec{u}, \  \text{where},$$
$$A = \begin{bmatrix} 0 & 0 & 1 & 0 \\ 0 & 0 & 0 & 1 \\ 3n^2 & 0 & 0 & 2n \\ 0 & 0 & -2n & 0 \end{bmatrix}, 
B = \begin{bmatrix} 0 & 0 \\ 0 & 0 \\ \frac{1}{m_c} & 0 \\ 0 & \frac{1}{m_c} \end{bmatrix},
\vec{u} = \begin{bmatrix} F_x \\ F_y \end{bmatrix}.$$

%%%%%%%%%%%%%%%%%%%%%%%%%%%
\subsection{Hybrid controller model}
\label{sec:hybrid-control}

Complete hybrid automaton models for the system with additional documentation are available from~\href{https://wiki.illinois.edu/wiki/display/MitraResearch/Autonomous+Satellite+System+Verification}{from this link}\footnote{https://tinyurl.com/verifysat}. 
Varying ranges of the relative distance between the spacecraft give rise to different constraints and requirements, and therefore, require separate controllers. We present a two-stage hybrid controller for achieving the rendezvous maneuver\footnote{The rendezvous mission presented in this paper is a subset of the four-stage problem presented in~\cite{erwin-cdc16}. Our two stages of rendezvous  are almost identical to ``Phase 2'' and ``Phase 3'' in~\cite{erwin-cdc16}.}. We refer to these discrete stages as \emph{modes}. 
Each discrete mode has an \emph{invariant} which specifies the conditions under which the system may operate in that mode, which we will first describe in words.
%Each mode is distinguished by an \emph{invariant} set of states for which the mode is valid, and we will first describe these invariants in words.

\begin{description}

\item{{\em Mode 1\/}} or Proximity Operations A (ProxA): the chaser is attempting to rendezvous and its separation distance ($\rho=\sqrt{x^2+y^2}$) from the target is in the range 100-1000m.
%when the separation distance ($\rho=\sqrt{x^2+y^2}$) between the two spacecraft is within 100-1000m, the chaser must maneuver towards the line-of-sight (LOS) region (see Figure \ref{}).

\item{{\em Mode 2}} or Proximity Operations B (ProxB): the chaser is attempting to rendezvous and its separation distance is less than 100m.
%In this mode, the chaser must remain inside the LOS region.

\item{{\em Mode 3}} or Passive mode: the chaser is no longer attempting to rendezvous and is not using its thrusters, regardless of its separation distance. The system may transition to the Passive mode as a result of a failure or loss of power.
\end{description}

The state of the overall hybrid system is defined by the mode and the valuations of a set of continuous variables: relative position $x$, $y$,
	thrusts $F_x$, $F_y$, and a global timer $\mathit{clock}$.	
There are two timing parameters of the model $t_1$ and $t_2$ that specify the time interval over which the chaser spacecraft may enter the Passive mode.
% Sayan: The following is a specific modeling choice and we will get to this when we describe the transitions in more detail
%A transition to the passive mode is governed by an interval of time ($[t_1,t_2]$), arbitrarily determined by the user. 
When the system is in a particular mode, the continuous variables $(x,y)$ evolve according to the (linear or nonlinear) differential equations of the previous section. The thrust inputs $F_x$ and $F_y$ are computed according the full-state feedback controller designed in Section~\ref{sec:lqr}. 

\begin{figure*}[t!]
\centering
\subfloat[]{
	\label{hyMod}
	\begin{tikzpicture}[scale=0.45]
	\tikzstyle{every node}=[font=\scriptsize, circle, font=\scriptsize, minimum size = 1.6cm,text width=2cm]
	\draw (-7,0) node[fill=blue!20] (p1) {\center{\vspace{-0.8cm}ProxA: $$\dot{\bar{x}} = (A-BK_1)\bar{x}$$}};
	\draw (0,0) node[fill=green!20] (p2) {\center{\vspace{-0.8cm}ProxB: $$\dot{\bar{x}} = (A-BK_2)\bar{x}$$}};
	\draw (7,0) node[fill=gray!20] (p3) {\center{\vspace{-0.8cm}Passive:$$\dot{\bar{x}} = A\bar{x}$$}};
	\tikzstyle{every node}=[font=\scriptsize,fill=none]
	\draw [ shorten >=1pt,->] (-7,4) to  node[left] {$\bar{x}_0$} (p1);
	\draw [ shorten >=1pt,->] (p1) to  [out=45,in=135] node[above] {$\rho\leq \rho_t$} (p2);
	\draw [ shorten >=1pt,->] (p2) to  [out=-135,in=-45]  node[above=4pt] {$\rho \geq \rho_t$} (p1);
	\draw [ shorten >=1pt,->] (p1) to  [out=-50,in=-140] node[above=1pt] {$\mathit{clock} \in [t_1,t_2]$} (p3);
	\draw [ shorten >=1pt,->] (p2) to  [out=45,in=135] node[above] {$\mathit{clock} \in [t_1,t_2]$} (p3);
	\end{tikzpicture}}
%	\caption{Hybrid model structure for spacecraft rendezvous. The flow equations shown here are linear. Replacing them with the nonlinear dynamics~(\ref{eq:nonlin1}) for each mode, would give a hybrid nonlinear model.}
%	\label{hyMod}
\subfloat[]{
	\label{rhofig}
	\hspace{5mm}
	\begin{tikzpicture}[scale=0.7]
	% ProxA region
	\fill [blue!20] (-2.4,-2.4) rectangle (2.4,2.4);
	% ProxB region
	\fill [green!20] (0,0) circle (2cm);
	% Axes
	\draw [ thick,->] (-2.5,0) to  node[right=2cm] {$\*i$} (2.5,0);
	\draw [ thick,->] (0,-2.5) to  node[above=2cm,right] {$\*j$} (0,2.5);
	% Rho
	\draw [shorten >=1pt,->] (0,0) to node[above,right]{$\rho_t = 100$m} (1.41,1.41);
	% Octogon
	\draw[thick] (0.83,2) -- (-0.83,2);
	\draw[thick] (0.83,2) -- (2,0.83);
	\draw[thick] (2,0.83) -- (2,-0.83);
	\draw[thick] (2,-0.83) -- (0.83,-2);
	\draw[thick] (-0.83,-2) -- (0.83,-2);
	\draw[thick] (-0.83,-2) -- (-2,-0.83);
	\draw[thick] (-2,-0.83) -- (-2,0.83);
	\draw[thick] (-2,0.83) -- (-0.83,2);
	\end{tikzpicture}}
%	\caption{The blue region represents a portion of ProxA's invariant and the green region, ProxB's invariant. The circumference of ProxB's invariant indicates a transition guard from ProxA to ProxB (i.e. when $\rho = 100m$.) The octagon circumscribing the circle is how the invariant and guard are approximated and modeled for the verification tools.}
%	\label{rhofig}
\caption{(a) Hybrid model for spacecraft rendezvous, with linear flow equations shown. The invariants in ProxA and ProxB are defined exclusively by the chaser's position, as shown by corresponding colors in the plane of motion in (b). The transition guards between ProxA and ProxB align exactly with their invariant sets, resulting in urgent transitions. The invariant for Passive mode is $clock > t_1$, irrespective of position. A transition to Passive occurs sometime within an interval of time, and hence is nondeterministic. In (b), the octagon represents how the invariants/guards are approximated and modeled in the verification tools.}
\end{figure*}
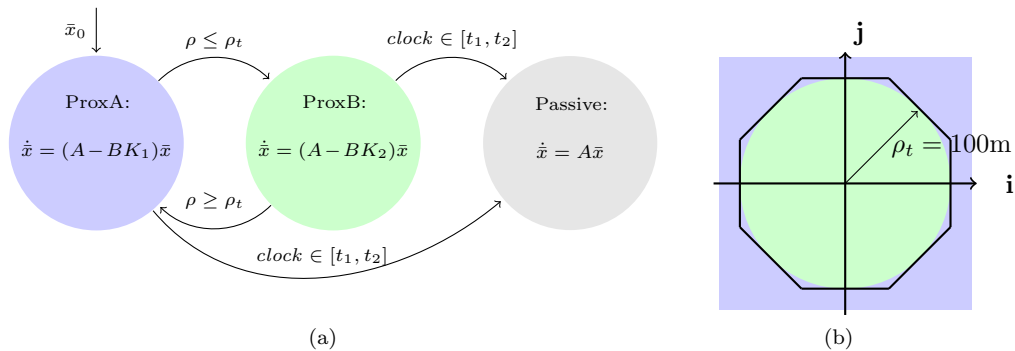

We refer to the time elapsed in the mission with the variable $clock$ but do not consider it an explicit state variable. The invariants in each mode can be more precisely described as $\rho \geq 100$ and $\mathit{clock} \leq t_2$ for mode 1, $\rho \leq 100$ and $\mathit{clock} \leq t_2$ for mode 2, and $\mathit{clock} \geq t_1$ for mode 3. 
A transition from one mode to another is described by a \emph{guard}. When the state satisfies the guard condition, the system \emph{may} take the transition. 
If a transition is required to occur as soon as possible, this is a called an \emph{urgent transition}. In this system, the distance-based transitions between modes 1 and 2 are urgent. However, the transitions to mode 3 (Passive mode) are not urgent. There is an interval of time, $\mathit{clock} \in[t_1,t_2]$, within which the chaser could nondeterministically transition to the Passive mode. Roughly, a larger $[t_1, t_2]$ interval implies a bigger passive-safety envelope for the mission. These transitions to the Passive mode make the system nondeterministic. Indeed, for some choices of this interval, it is possible for the hybrid system to occupy any one of the three modes at a given time.

%% Moved to Verification Approaches section
%The hybrid system verification tools like  C2E2~\cite{DuggiralaMVP15,FanQM0D16:CAV} and SpaceEx~\cite{DBLP:conf/cav/FrehseGDCRLRGDM11} require that the initial set of states are bounded and convex and that guards and unsafe sets of states are presented at polyhedra. Therefore, the quadratic constraints imposed by  $\rho$ are  approximated using polygons (See Figure \ref{rhofig}).
%Additionally the invariant of mode 1 cannot be exactly modeled for SpaceEx since it is not convex. Instead, we only specify one edge of the polygon. 

\begin{figure*}
\centering
\begin{minipage}[t]{.47\textwidth}
	\centering
	\includegraphics[width=.65\textwidth]{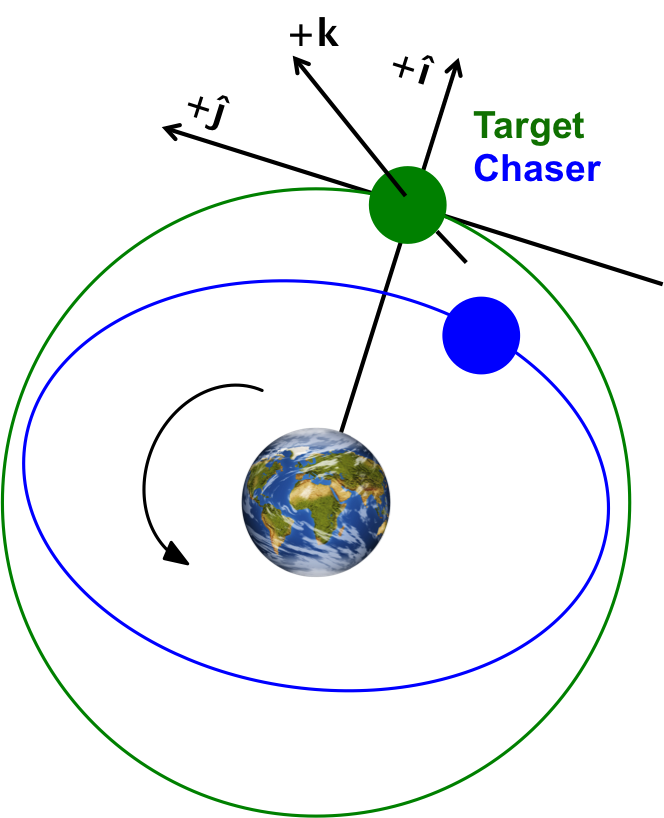}
	\caption{Hill's relative coordinate frame. The chaser's relative position vector is $x\hat{\*i}+y\hat{\*j}$.}
	\label{Hill}
\end{minipage}\qquad
\begin{minipage}[t]{.47\textwidth}
	\centering
	\begin{tikzpicture}[scale=0.8]
	% ProxA region
	\fill [blue!20] (-2.5,-2.5) rectangle (2.5,2.5);
%	\fill [blue!20] (0,0) circle (2.5cm);
	% ProxB region
	\fill [green!20] (0,0) circle (1.5cm);
	% LOS 
	\fill [red!60] (0,0) -- (-1.3,0.75) arc (150:210:1.5) -- cycle;
	% Axes
	\draw [ thick,->] (-3,0) to  node[right=2.5cm] {$\*i$} (3,0);
	\draw [ thick,->] (0,-3) to  node[above=2.5cm,right] {$\*j$} (0,3);
	
	\draw [shorten >=1pt,->] (0,0) to node[above]{$\rho$} (1,1.33);
	\draw (0,0) circle (1.67cm);
	\draw (-0.87,0.5) arc (150:210:1) node[right] {$60^{\circ}$};
	\end{tikzpicture}
	\caption{The hybrid model (see Figure~\ref{hyMod}) captures the chaser's motion in ProxA (blue) and ProxB (green), and we use verification tools to show that whenever $\rho \in$ ProxB, the chaser does not leave the LOS region (red).}
	\label{LOSfig}
\end{minipage}
\end{figure*}

%%%%%%%%%%%%%%%%%%%%%%%%%%%
\subsection{Linear Quadratic Control}
\label{sec:lqr}
We have developed a full-state feedback controller, namely, a Linear Quadratic Regulator (LQR), to drive the chaser towards the target's position. 
%We assume we have a perfect state observer, so that we have access to current state information. Though we do not strictly model this observer, we know it is achievable in theory because this system is fully observable. 
Closed-loop feedback is desirable because the system can measure and adjust for errors, and ultimately guarantee liveness (i.e. eventually the target will be reached). LQR is specifically chosen because it is constructed by minimizing a quadratic cost function, which we can choose so as to roughly satisfy our safety constraints. 
LQR is only applicable to linear systems, so we design the control for the linearized model in~\eqref{eq:lin1}, but we will use the same control with nonlinear dynamics~\eqref{eq:nonlin1} when applying verification tools.

The form of an LQR solution is: $\vec{u} = -K\vec{x}$, where $K \in \mathbb{R}^{4\times2}$ is a constant matrix and $\vec{u} = [\frac{F_x}{m_c}, \frac{F_y}{m_c}]^T$. The $K$ matrix is found by minimizing the following cost function with respect to $\vec{u}$: $\int_{0}^{\infty} (\vec{x}^{T}Q\vec{x}+\vec{u}^{T}R\vec{u}) dt $, where $Q$ and $R$ are positive definite matrices.

Given the form of the control law, we update the definition of the %\linfour 
model given in~\eqref{eq:lin1} to the following:

\begin{align}
\begin{split}
\ddot{x} &= \left(3n^2-\frac{k_{11}}{m_c}\right) x - \frac{k_{12}}{m_c}y - \frac{k_{13}}{m_c}\dot{x} + \left(2n-\frac{k_{14}}{m_c}\right)\dot{y},\\
\ddot{y} &= - \frac{k_{21}}{m_c}x -  \frac{k_{22}}{m_c}y - \left(2n+ \frac{k_{23}}{m_c}\right)\dot{x} - \frac{k_{24}}{m_c}\dot{y}.
\label{eq:cl}
\end{split}
\end{align}

These equations are the expanded version of the closed-loop form shown in Figure~\ref{hyMod}. 
Later, we will discuss why we distinguish between the dependence of~\eqref{eq:lin1} on $\vec{x}$ and $\vec{u}$ and~\eqref{eq:cl} on only $\vec{x}$.

Bryson's method \cite{brysonrule} is used to help determine an appropriate cost function. Begin with $Q$ and $R$ as diagonal matrices, and choose their values so as to normalize each of the state and input variables. In other words, choose the diagonal elements so that $Q_{ii} = \frac{1}{max(x_i^2)}$ and $R_{ii} = \frac{1}{max(u_i^2)}$. Here, the denominators refer to the largest \emph{desired} value of each variable, which will be determined by the safety constraints and mode invariants. While the LQR gains are obtained with our constraints in mind, the resulting controller does not guarantee these constraints are never violated. This is why further verification is still required.
This design process is repeated for modes 1 and 2, and the result is two distinct LQR controllers for each of these modes in our hybrid system.

%%%%%%%%%%%%%%%%%%%%%%%%%%%
\subsection{Constraints and safety requirements}
\label{sec:safeSets}
%The goal of the verification process is to overapproximate the reachable states of the system and check for intersection with unsafe sets of states. If no intersection is found, then the system is guaranteed to be safe. %First we will present the safe region of states, followed by the model of the complementary unsafe sets.
In this section, we enumerate the properties that define a safe and successful mission, and how they are modeled for verification tools. 
%C2E2 and SpaceEx. 

	\begin{description}
		\item[ Thrust constraints]
During the rendezvous stages (ProxA and ProxB), the thrusters cannot provide more than $10 N$ of force in any single direction, therefore, we have the constraints:
 \[
 |F_x|, |F_y| \leq 10. %\implies |u_x|,|u_y| \leq \frac{10}{m_c} = 0.02.
 \] 
\item [LOS cone and proximity]
During close-range rendezvous ProxB, the chaser must remain within a line-of-sight (LOS) cone (see Figure \ref{LOSfig}), and its total velocity must remain under $5$cm/s, so $\sqrt{\dot{x}^2+\dot{y}^2} \leq 5$cm/s. The total velocity constraints cannot be exactly modeled using linear constraints, and a polytopic approximation over $\dot{x},\dot{y}$ is used. This is done in the same way as $\sqrt{x^2+y^2}\leq \rho_t$ is approximated (see Figure~\ref{rhofig}).
\item[Separation]
During the Passive mode, the chaser must avoid collision with the target, which is theoretically a point mass at the origin. Even in a theoretical model, a small ball or box should be used to bound this point to account for limitations in numerical precision. In reality, the target satellite's dimensions may range from the order of $1$m %(Sputnik 1) 
to $100$m, %(International Space Station)
so the size of this bounding box will vary depending on the situation. We use a box with a $0.1$m circumradius. 

\end{description}

\section{Verification approaches}
In this section, we briefly discuss our experience in using hybrid system verification tools. 
{\bf SpaceEx\/}~\cite{DBLP:conf/cav/FrehseGDCRLRGDM11}, is a well-established reachability analysis tool for linear and affine hybrid systems. It implements the support function-based reachability algorithm, includes the PHAVer algorithm for rectangular dynamics~\cite{Frehse:hscc05}, and also a simulator for nonlinear models. 
The support function representation of sets is amenable to effective computation of convex hulls, linear transforms, Minkowski sums, etc.---operations that are necessary for safety verification. 
%SpaceEx also has a model editor for constructing networks of hybrid automata. 

{\bf C2E2\/}~\cite{DuggiralaMVP15,FanQM0D16:CAV} is a simulation-driven bounded verification tool for nonlinear hybrid models. 
The core algorithm of C2E2 relies on computing reachset over-approximations from validated numerical simulations and what are called {\em discrepancy functions.\/} A discrepancy function for a model bounds the sensitivity of the trajectories of the hybrid system to changes in initial states and inputs. Candidate discrepancy functions can be obtained using a global Lipschitz or using a matrix norm for linear systems.
However, typically these approaches give discrepancy functions that blow-up exponentially with time, and therefore, are not useful for verifying problems with long time horizons.
The automatic on-the-fly approach implemented in~\cite{FanQM0D16:CAV} uses bounds on the Jaobian matrix of the system to get tighter local discrepancy functions and it has been used to verify several benchmark problems. Recently the tool has been extended to handle nonlinear models with dynamics with exponential and trigonometric functions. 

For a (possibly nonlinear) mode with $\dot x = f(x(t))$, the discrepancy computed by the algorithm of~\cite{FanQM0D16:CAV} uses the Jacobian matrix $J(x)$ of $f(x)$ and the condition number  of $J(x_0)$ evaluated at certain points $x_0$ in the state space. For ill-conditioned matrices, such as what we have in the passive mode, (the $A$-matrix representation of \eqref{eq:lin1}), the over-approximation error may still blow-up.
Ill-conditioned systems may not only arise from passive dynamics but also from extremely large and small coefficients appearing together in $J(x_0)$.

In order to address this problem, we have created a MATLAB implementation of C2E2's verification algorithm ({\bf\sdvtool{}}) for linear models.
Unlike C2E2, \sdvtool{} \ does not rely on discrepancy, but instead  computes the reachable states under a given linear mode directly.
The particular algorithm implemented is the one presented in~\cite{Duggirala2016}:
% 
%\subsection{Improved Algorithm for Linear Models}
%In this section, we present an alternate algorithm for obtaining overapproximated reachsets derived from exact reachsets. One of the challenges of obtaining exact reachsets is how to represent the reachsets. SpaceEx uses support functions and C2E2 uses convex polyhedra. The latter is more straightforward to work with, so we use a different approach from SpaceEx. The new approach is found in \cite{}. %We implement this approach in Matlab first and plan to implement it in C2E2 as well. 
%
For an $n$-dimensional system, $n+1$ simulations are performed. From these simulations, special sets called \emph{generalized star sets}, are generated to represent the exact reachsets. 
For our purposes, a generalized star set is represented by a pair $\langle x_0,V \rangle$, where 
%\begin{enumerate}
%	\item  
	$x_0 \in \mathbb{R}^n$ is the center state and  
%	\item  
	$V=\{v_1,...,v_n \} \subseteq \mathbb{R}^n$ is a standard basis (not necessarily unit vectors), and 
	%, and 
%	\item  $P: \mathbb{R}^n\rightarrow \{\bot,\top\}$ is a predicate of the following form:
%	$$ P(\bar{\alpha}) = 
%	\begin{cases} 
%	\hfill \top    \hfill & \text{ if $|\alpha_i| \leq 1$, $\forall i=1,2,...,n$} \\
%	\hfill \bot    \hfill & \text{otherwise}. \\
%	\end{cases} $$
%\end{enumerate} 
the set defined by $\langle x_0,V \rangle$ is 
 $$\{x \in \mathbb{R}^n \ | \  \exists \alpha_1, \ldots, \alpha_n \in [-1,1],  x = x_0 + \sum_{i=1}^n \alpha_i v_i \}.$$
%
%
%We restrict the initial set to a hyperrectangle and use a standard basis $V=\{v_1,...,v_n \} \subseteq \mathbb{R}^n$ for the initial set. The vectors in $V$ are not necessarily unit vectors but are scaled to the radius of the initial set in each dimension. Then,
%
As reachsets are calculated for time steps, $x_0$ and $V$ are transformed.
When the reachtube from a given mode intersects the guards for a transition, the star sets are aggregated and over-approximated with hyperrectangles. 
%
% In order to avoid introducing new data structures into C2E2, we represent a reachset with a hyperrectangle that overapproximates the star set.  
 If $R_i ^*= <x(t_i), V_i>$ is the star set reachset obtained at time $t_i$, then the hyperrectangular reachset is: 
\begin{equation*}
\begin{aligned}
R_i = \{x \ | \ x\leq x(t_i)+\sum_{j=1}^n max(-v_j,v_j) \text{ }\\ \text{and } x\geq x(t_i)+\sum_{j=1}^n min(-v_j,v_j) \}.
\end{aligned}
\end{equation*}
%This implementation is used for verifying the linear models \linfour \ and \linsix.

C2E2 and \sdvtool{} currently accumulates all the reachable sets in ProxA and ProxB that \emph{may} transition to Passive, and uses their convex hull to begin reachset computations under the Passive mode. It follows that if the time interval during which a transition may occur is large, then the initial set of states under the Passive mode is large, making it very difficult to prove safety. One solution is to allow partitioning and refinement of the initial passive mode set. Since this is not currently implemented in C2E2 or \sdvtool{}, we restrict our experiments to transition interval lengths of 5 minutes or less. 
For example, checking if the system is safe for a transition $clock\in [50, 200\text{ min}]$ could be achieved by running several experiments with small subintervals that cover the original interval.

\section{Verification results}
\label{sec:results}

%\sayan{The figures should be smaller; say, 0.45textwidth; the fonts in the labels have to be increased accordingly. Also the spaceex and MATLAB figures for the same computation should be side-by-side for comparison.}

In this section, we discuss and compare verification results from SpaceEx, C2E2, and our implementation of \sdvtool{}. Based on these results, we reach the broad conclusion that with some manual tweaks, the current hybrid system verification tools are indeed capable of analyzing realistic system-level properties of autonomous spacecraft maneuvers. 

In the following presentation, we pick arbitrary model parameters, but to a large extent our results are robust with respect to parameter variations.
That is, the parameters can be tuned to the specific requirements of a real mission. 

For subsequent discussion, we label our models as follows: \linfour{} denotes the equations in \eqref{eq:cl}, \nonlinfour{} denotes the equations in \eqref{eq:nonlin1} with the same controller as \linfour{} substituted into $F_x,F_y$, and \linsix{} denotes a model that will soon be introduced to account for explicit thrust values. 

%We show an example of how to design and analyze a more complex model of the mission, and we will discuss other ways to increase complexity to tailor to the user's needs. 

\begin{figure*}[t!]
	\centering
	\subfloat[]{\label{mat1:a}\includegraphics[width=.37\textwidth]{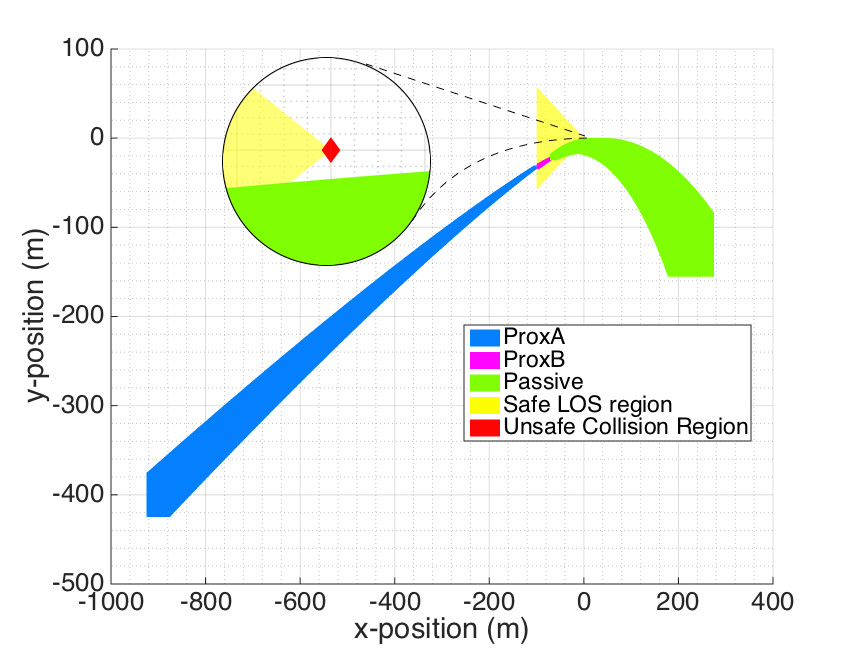}}%
	\subfloat[]{\label{space1:a}\includegraphics[width=.31\textwidth]{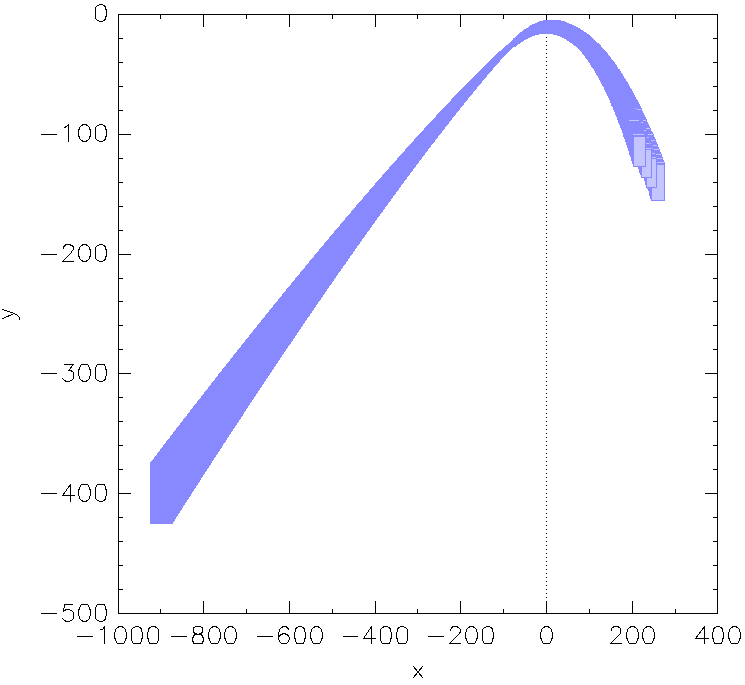}}%
	\subfloat[]{\label{c2e2:a}\includegraphics[width=.36\textwidth]{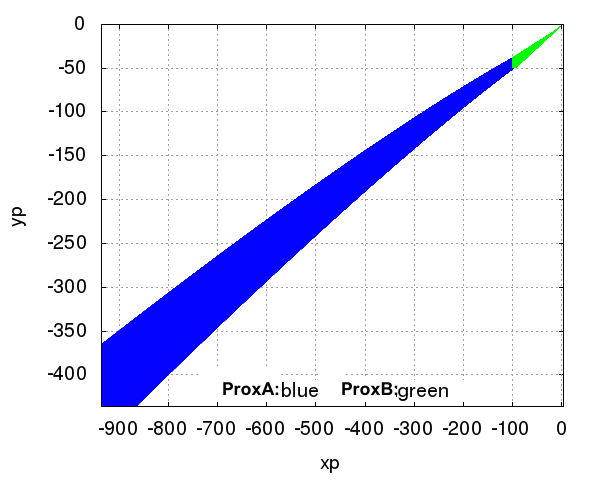}}\\
	\subfloat[]{\label{mat1:b}\includegraphics[width=.37\textwidth]{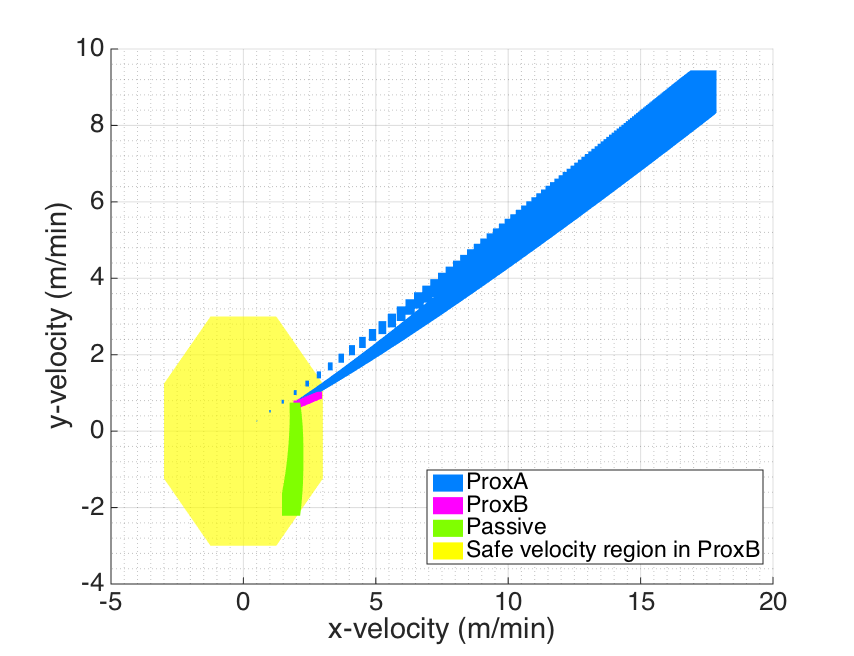}} %
	\subfloat[]{\label{space1:b}\includegraphics[width=.31\textwidth]{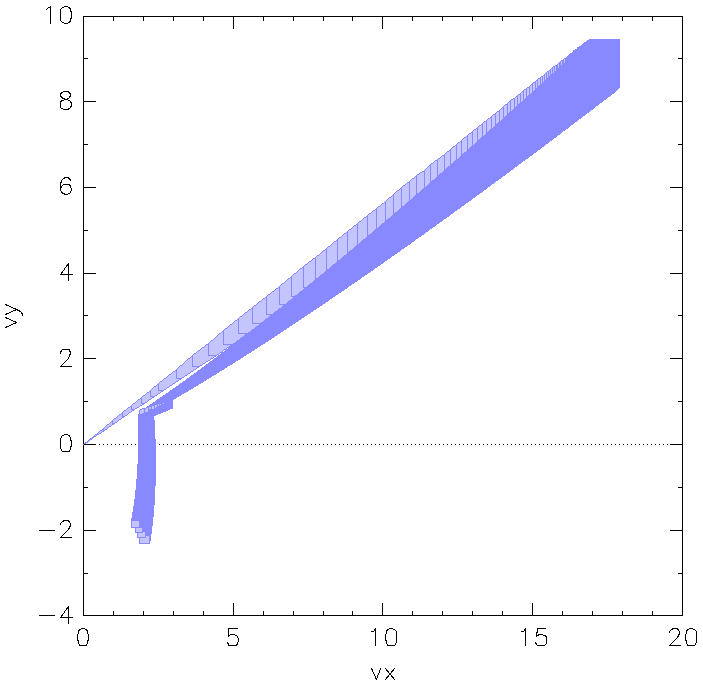}}%
	\subfloat[]{\label{mat3}\includegraphics[width=.36\textwidth]{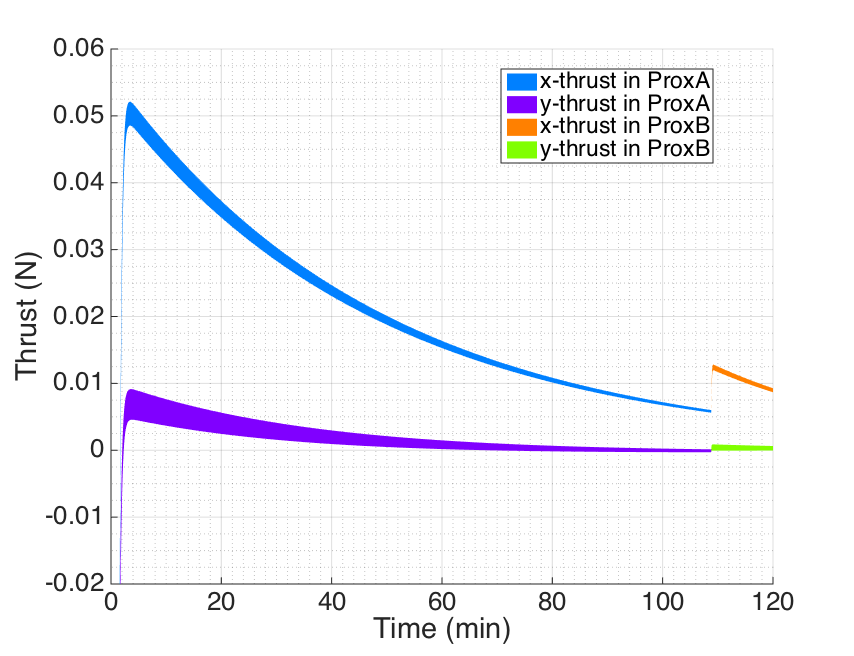}}%
	\caption{Examples of various generated reachsets. Reachable positions using \linfour{} in \sdvtool{} shown in (a) and SpaceEx in (b). Reachable velocities using \linfour{} in \sdvtool{} shown in (d) and SpaceEx in (e). Reachable positions of \nonlinfour{} without Passive mode in C2E2 shown in (c). Reachable thrusts using \linsix{} in \sdvtool{} is shown in (f).}
	\label{results}
\end{figure*}

\subsection{Hybrid safety proofs}
Figure~\ref{results} shows the typical reachset computations obtained from \sdvtool{}, C2E2, and SpaceEx on the \linfour, \linsix, and \nonlinfour \  models. These computations also establish the safety of the corresponding systems with respect to the requirements in Section~\ref{sec:safeSets}. 
Overall, the plots show that the reachsets from the different tools are qualitatively similar. 
From the more detailed MatLab plots we can check that no part of the reachable sets intersect with unsafe regions. It is clear from the zoomed in portion of Figure~\ref{mat1:a} that a reasonably larger collision region would violate safety. 

In C2E2 and SpaceEx, each safety property is loaded and checked individually. In C2E2, the running time for a single property for the nonlinear model \nonlinfour{} is in the neighborhood of 5-10 minutes; in SpaceEx, the running time for a single property is on the order of a few seconds. \sdvtool{} checks all (12) properties simultaneously and the running time varies from around 30 seconds to 10 minutes. We do not compare absolute running times in further detail in this paper as each of the tools have different semi-automatic workflows and require widely different execution environments.

%The visualization of support functions in SpaceEx can be approximated by boxes or octagons, and we chose boxes to better match the hyperrectangular sets from Matlab. A Matlab reachset is represented by a single hyperrectangle, whereas SpaceEx uses the union of multiple boxes to represent a reachset, reducing overapproximation error.

% Sayan, I don't think it makes sense to compare runtimes because in SpaceEx, I have to load a different configuration file for each unsafe property (12 total, +4 when we add in thrust constraints). Whereas matlab just runs once to check for all the properties. I COULD sum up the independent runtimes for SpaceEx. But I think it's a waste of time to do this right now. 

%\begin{description}
\paragraph{Time horizon} 
Timing is obviously critical for space applications to ensure there is sufficient fuel, but with over-approximated reachability, we can only guarantee the mission is completed within a  time upper-bound. This upper-bound obtained from reachability analysis may differ significantly from the actual mission time.
%
%We treat the mission's total time duration as a soft constraint. Alternatively,  we  set a goal set of states (e.g. the chaser arrives within $10m$ of the target under $5$ hours). 
%Then we would add an unsafe set: chaser may be farther than $10m$ from target after 5 hours. This is an example of checking a strict mission completion constraint. 
% SM: best not to talk about stuff we could/will do. 
%
 Therefore, we do not impose any strict completion requirements. Instead, we choose a time horizon that is representative of what we might expect in practice, and focus on observing what behaviors are possible within these limits. Typically, Proximity A operations take 1-5 orbits (at under 4 hours an orbit) and proximity B operations take 45-90 minutes~\cite{wertz:phases}. We choose a sum of approximately 4.5 hours to be our time horizon.
% An orbit takes $\frac{1}{n}$ time, which is a little under 4 hours for this scenario, so we choose a sum of approximately 4.5 hours to be our time horizon.

\paragraph{Initial states} We calculate a set  of initial states assuming that the chaser spacecraft is performing the encompassing mission from~\cite{erwin-cdc16}. 
%Then, the chaser begins at a point along the $-\*j$-axis, with a separation distance significantly greater than that of mode ProxA's invariant. The chaser navigates through the $-\*i, -\*j$ quadrant to get from this initial location to the LOS region. We choose initial states for ProxA that are in this quadrant. 
% SM: these details are not so important.
%For an arbitrary first initial state, we choose $\vec{x}_0 = [-900m,-400m,0m/s,0m/s]^T$. Its separation distance is close to one that would trigger a transition to mode ProxA, and it is positioned along the same direction as the edge of the LOS cone. Later we will also show coarse results using a range of initial states in the same quadrant with similar separation distances.
%
%For this first scenario, 
We choose an initial set radius of $[25m,25m,0,0]$
around the point $\vec{x}_0 = [-900m,-400m,0m/s,0m/s]^T$. This can be interpreted as uncertainty in the chaser's initial position, typically due to loss of precision from sensors and computations, or it can be used to explore multiple initial states of interest. We have successfully verified scenarios with  uncertainty in the velocity dimensions as well.
% We omit these results to simplify the presentation of our results.
% SM: Is there any significant 

\paragraph{Unsafe sets} For SpaceEx, C2E2, and \sdvtool{}, we model the safety requirements as a collection of linear inequalities. 
The LOS cone is approximated with a triangle, so we check three properties to prove the system remains within LOS constraints, and so on. 
%The collision region is approximated with a square, resulting in four properties. Total velocity is approximated with an octagon and requires eight properties. 
Max thrust is effectively a one-dimensional constraint, a nonconvex interval, so two properties will capture the unsafe set for each thrust input (one along $x$-direction, one along $y$-direction). But in order to treat it this way, we must introduce extra variables $u_x,u_y$ to explicitly track the thruster values. These extra variables are the difference between \linfour{} and \linsix{}.

\paragraph{Passive transition time} The interval of time during which a transition to the passive mode may occur is trivially bounded by the  mission time horizon. For this first example, we choose a small interval at $[120,125
\text{min}]$. This ensures that the chaser will operate in mode ProxB before transitioning to the passive mode.

\subsection{Adding thrust constraints}
\label{sec:6dim}
\begin{figure*}
\centering
\begin{minipage}[t]{.47\textwidth}
	\centering
	\includegraphics[width=\textwidth]{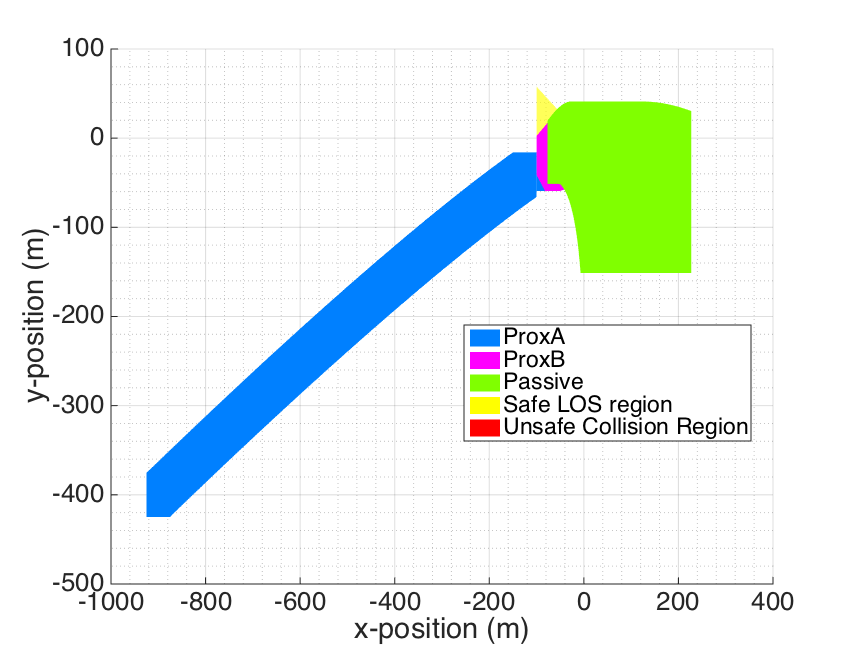}
	\caption{Coarse overapproximation of the reachable positions, when using the 6-dimensional \linsix{} model from Equation~\ref{eq:6dim1}.}
	\label{mat2}
\end{minipage}\qquad
\begin{minipage}[t]{.47\textwidth}
	\centering
	\includegraphics[width=\textwidth]{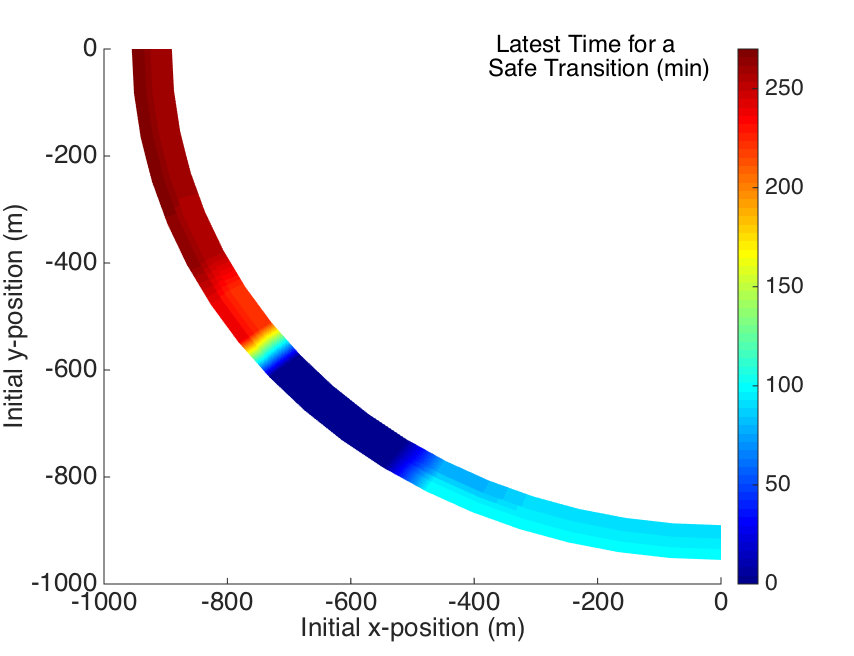}
	\caption{Initial positions (with zero initial velocities) of \linfour{} that have been verified to be safe. They are safe for Passive transition times up to the time shown by the color map.}
	\label{param1}
\end{minipage}
\end{figure*}

 In Section~\ref{sec:safeSets}, we described a constraint on thrust that mimics the physical limitations of our spacecraft. We now set up the 6-dimensional model \linsix \ so that we can verify this additional requirement. We introduce $u_x,u_y$ as explicit state variables, and solve for their differential equations to obtain:
\begin{align}
\begin{split}
\dot{u}_x &= k_{11}\dot{x} + k_{12}\dot{y} + k_{13}\ddot{x} + k_{14}\ddot{y},\\
\dot{u}_y &= k_{21}\dot{x} + k_{22}\dot{y} + k_{23}\ddot{x} + k_{24}\ddot{y}.
\end{split}
\label{eq:thrustode}
\end{align}
There are two equivalent numerical models that will produce different over-approximated reachsets. The first model consists of \eqref{eq:thrustode} and the following:
\begin{align}
\begin{split}
\ddot{x} &= 3n^2 x + 2n\dot{y} - \frac{u_x}{m_c},\\
\ddot{y} &= - 2n\dot{x} - \frac{u_y}{m_c}.
\end{split}
\label{eq:6dim1}
\end{align}
Here $\ddot{x}$ and $\ddot{y}$ account for the effects of thrust inputs by explicitly adding $u_x,u_y$. Since each dimension is over-approximated in the reachset computation and $u_x,u_y$ are functions of position and velocity, the computation for subsequent reachable sets of position and velocity have even more uncertainty. 
Roughly, $u_x,u_y$ act as filters for $x,y,\dot{x},\dot{y}$, adding distortion and introducing more uncertainty. 
Figure~\ref{mat2} shows the effects of these compounding errors. The overarching verification algorithm will partition the initial set to reduce errors stemming from the data structure, but it will have to do this numerous times and may time out in practice.

The second  6-dimensional model (a variant on \linsix) consists of~\eqref{eq:cl} and \eqref{eq:thrustode}. In this case, $\ddot{x}$ and $\ddot{y}$ implicitly calculate the thrust, and $u_x,u_y$ are independent ``tracking'' variables. The calculations for $\ddot{x}$ and $\ddot{y}$ are equivalent to those in \eqref{eq:6dim1}, but they bypass the ``filter'' when constructing reachsets. The results are identical to those shown in Figures~\ref{mat1:a}-\ref{mat1:b}, with the addition of reachable sets of thrusts shown in Figure~\ref{mat3}.

%In either 6-dimensional models, it is the user's responsibility to properly input the initial set of states for $u_x,u_y$. They should, at a minimum, encompass the values corresponding to the set of initial positions and velocities. This is necessary for the \eqref{eq:cl}-\eqref{eq:thrustode} model, where $u_x,u_y$ evolve according to a rate determined by the decoupled position and velocity variables. This evolution will not change between different initial sets of $u_x,u_y$, but the results are meaningless if the explicit and implicit initial values of $u_x,u_y$ do not match.
 Once again, our choice of data structure introduces some uncertainty to the explicit representation of the initial set of $u_x,u_y$ values. This is propagated throughout the analysis.
 %, so it is recommended that the model with least-dimensionality is used with the verification tools. 
 %In other words, if \linfour suffices for checking the LOS constraint, then it is unnecessary do not use \linsix for checking this property.
%
We use the \eqref{eq:cl}-\eqref{eq:thrustode} model to obtain safe thrusting results from \sdvtool{}. The fine reachable sets in Figure~\ref{mat3} show that the LQR controller operates well within the thrust constraints ($|u_x|,|u_y| \leq 10N$).

\subsection{Robustness of verification}
To  demonstrate the robustness of the verification approaches (and the designed controller), we performed several experiments varying the initial set and  passive transition times with the \linfour \ model. 

The scenarios that guarantee a safe mission are  summarized in Figure~\ref{param1}. 
Roughly, choosing an initial position or subset of positions within the shaded region will result in a safe mission for a transition time or interval within $[0,T]$, where $T$ is the time corresponding to the color at that initial position(s). 
For these experiments, we consider initial separation between the chaser and the target to be near $1000m$, where this LQR control would start being used. We assume the initial chaser velocity to be zero. 
Generally, we can conclude that, the closer to the $x$-axis the chaser starts, the later the chaser may safely abort to the passive mode. 
On the other hand, the neighborhood of states along $\sim 230^{\circ}$ are not safe for a passive transition at any time. 
%\begin{figure}[h!]
%	\centering
%	\includegraphics[width=.52\textwidth]{fig2.png}
%	\caption{Ranges of initial states and transition times to Passive mode that result in a safe mission.}
%	\label{param1}
%\end{figure}

\section{Conclusions and future directions}
\label{sec:conclusion}

In this case study paper, we present a sequence of linear and nonlinear, nondeterministic benchmark models of autonomous rendezvous between spacecraft with several physical and geometric safety requirements. We designed an  LQR controller and verified its safety across the different models, a variety of initial conditions, parameter ranges, and using three different hybrid system verification approaches. The models and requirements are made available online.

This case study, and in particular the requirement for passive safety, has  shed light on the weakness of simulation-driven verification in handling ill-conditioned models.

The results provide a foundation for verifying  more sophisticated maneuvers in future autonomous space operations.
For example, we proposed a continuous full-state feedback controller, but it is also possible to consider a situation where full state measurement is not possible and a simple bang-bang controller is required. Control theory tells us that this system maintains marginal stability which implies that errors will never recede, so for reasonably-sized initial sets, the reachable sets may not satisfy tight constraints such as LOS. 

%Transitions in hybrid models are able to capture such nondeterminism. 
%Other nondeterministic transitions may also be introduced. In Section~\ref{sec:safeSets}, we described the approximation of the guard from ProxA to ProxB with an octagon. Specifically, we chose its radius to over-approximate the exact boundary. Say, we would like to choose a region that encompasses the exact guard boundary, in which a nondeterministic transition occurs. This region can be formulated using the intersection of overapproximated and underapproximated octagons. 

%\paragraph{Adding Constraints} 
%Other variations: 3 DOF model, discrete thrusting (more realistic), adding in attitude dynamics and constraints, different controller, other phases (docking, relocation)

\section*{Acknowledgments}
We are grateful for the support of Richard S. Erwin in navigating and modeling the problem presented in this paper, and for Yu Meng's support with the C2E2 experiments. We acknowledge the support of the Air Force Research Laboratory through the Space Scholars Program. 

\bibliographystyle{abbrv}
\bibliography{Chan_iccps,sayan1}

\end{document}